\documentclass[lettersize,journal]{IEEEtran}
\usepackage{amsmath,amsfonts}
\usepackage{gensymb}
\usepackage{algorithmic}
\usepackage{algorithm}
\usepackage{array}
\usepackage[caption=false,font=normalsize,labelfont=sf,textfont=sf]{subfig}
\usepackage{textcomp}
\usepackage{flushend}
\usepackage{stfloats}
\usepackage{url}
\usepackage{verbatim}
\usepackage{graphicx}
\usepackage{xcolor}
\usepackage{cite}
\usepackage[acronym]{glossaries}
\makeglossaries
\newacronym{ber}{BER}{bit error rate}
\newacronym{co}{CO}{carbon monoxide}
\newacronym{co2}{CO\textsubscript{2}}{carbon dioxide}
\newacronym{cplc}{CPLC}{contactless PLC}
\newacronym{emi}{EMI}{electromagnetic interference}
\newacronym{epr}{EPR}{ethylene propylene rubber}
\newacronym{esl}{ESL}{series inductance}
\newacronym{ghz}{GHz}{Gigahertz}
\newacronym{los}{LoS}{line of sight}
\newacronym{lti}{LTI}{linear time-invariant}
\newacronym{mv}{MV}{medium oltage}
\newacronym{mom}{MoM}{method of moments}
\newacronym{pcr}{PCR}{power cable radiation}
\newacronym{plc}{PLC}{power line communication}
\newacronym{rf}{RF}{radio frequency}
\newacronym{shf}{SHF}{super high frequency}
\newacronym{snr}{SNR}{Signal-to-Noise Ratio}
\newacronym{tte}{TTE}{through-the-earth communication}

\usepackage[justification=centering]{caption}
\begin{document}
\bstctlcite{IEEEexample:BSTcontrol} 

\title{ Power Cable Radiation: A Novel Approach to Underground Mining Connectivity}

\author {{Siphiwe Shandu}, {Thabiso Moropa}, and {Alain R. Ndjiongue}\thanks{The authors are with the School of Electrical and information Engineering, University of the Witwatersrand, 1 Jan Smuts Ave, Braamfontein, Johannesburg, 2017.}}


\maketitle

\begin{abstract}
This letter investigates contactless power line communications (CPLC) for underground mining by modeling power wires as long-wire antennas. A system-level framework is developed, comprising a cascade of RF and power line channels. The model accounts for multipath propagation, frequency-dependent attenuation, and Rician fading. Simulations from 1-20 GHz reveal that the length of the wire significantly affects radiation, directivity, and input impedance. The findings show that CPLC transmits electromagnetic waves without direct electrical contact, offering a robust, cost-effective solution that enhances mobility, reduces maintenance, and ensures compatibility with existing mining power infrastructure.
\end{abstract}


\section{Introduction}
\IEEEPARstart{U}{nderground} mining environments present unique challenges for communication systems due to dynamic conditions such as shifting topologies, high humidity (greater than 90\%), corrosive water, dust, and electromagnetic interference from heavy machinery, all of which degrade signal reliability \cite{5208737, misra2010underground}. In addition to these harsh physical conditions, the potential presence of explosive gases demands the use of robust and explosion-proof equipment to ensure safe operation \cite{5208737}. In these environments, effective communication is critical for safety, allowing emergency alerts and evacuation coordination, and for operational efficiency through real-time orchestration. Traditional wired systems, such as magnetophones, offer simplicity but lack mobility and require frequent maintenance due to cable wear \cite{5208737}. Radio-based systems, including  through-the-earth communication and wireless networks, improve mobility, but suffer from \gls{los} limitations and signal attenuation in complex mine topologies \cite{misra2010underground}. Other technologies, such as leaky feeders, are susceptible to cable damage and require power at the interfaces. A potential technology to be used in this environment is \gls{plc} that exploits power cables for data transmission, but is limited by the need for direct electrical connections \cite{debeer2016contactless}.

In this work, we present a novel \gls{cplc} model that leverages radiative coupling to transmit signals through electromagnetic fields, treating power cables as antennas \cite{debeer2016contactless, igboamalu2021contactless}. This approach eliminates direct electrical contact, enhancing mobility for equipment and personnel in mines where \gls{los} issues and attenuation are prevalent \cite{5208737, misra2010underground}. By leveraging existing power infrastructure, \gls{cplc} reduces deployment costs and can serve as an environmental sensor for monitoring gas levels, temperature, and pressure. More specifically, this letter develops a theoretical framework for \gls{cplc} by modeling power cables as long-wire antennas, applying antenna theory principles, and simulating their performance in a mining environment, considering carrier signals with frequencies from 1-20 GHz. We focus on a $\kappa$-wire system coupled to a \gls{rf} source. To this end, the primary contributions of this paper are: (\textit{i}) The evaluation of the radiation characteristics of an indoor long-wire antenna in a lossy environment; (\textit{ii}) An analysis of the directivity characteristics of an indoor long-wire antenna in a mining tunnel; (\textit{iii}) Modeling of an indoor power wire antenna at specific frequencies based on the gain and directivity profile; (\textit{iv}) We also characterize the three-phase power cable typically used in underground mining applications; (\textit{v}) Finally, we develop a system and channel model for indoor long-wire antenna and application to underground mining communications.

\section{System Model}
\begin{figure}
    \centering
    \includegraphics[width=0.76\linewidth]{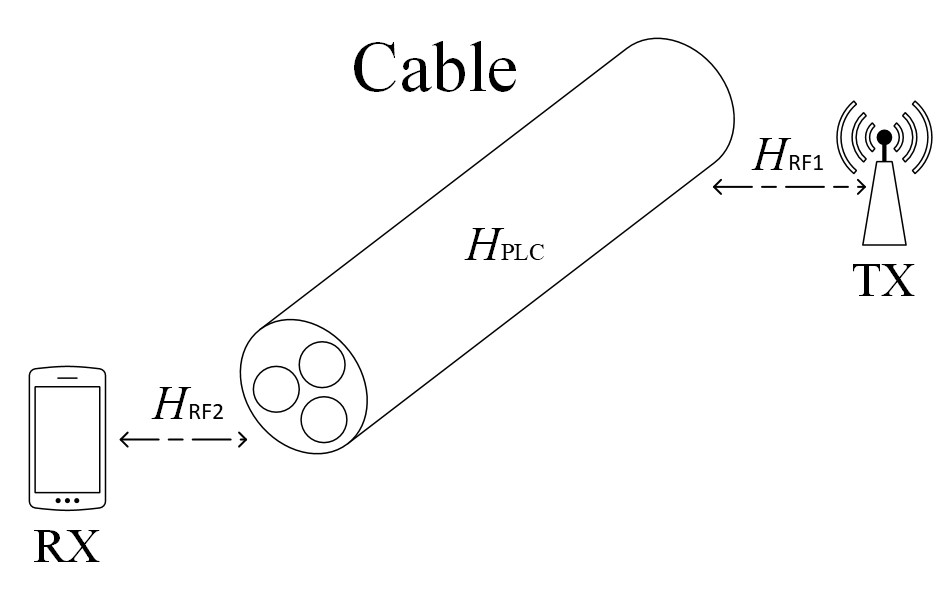}
    \caption{CPLC system diagram.}
    \label{fig:cplc_system}
\end{figure}

 In this section, we discuss the foundational principles of radiation, focusing on application to long-wire antennas. A single radiating conductor is assumed, neglecting the mutual coupling and combined radiation effects of multiple simultaneously active wires. According to Fig. \ref{fig:cplc_system}, the overall transfer function can be expressed as \(H_{\text{CPLC}} = H_{\text{RF1}} H_{\text{PLC}} H_{\text{RF2}}\), where $H\textsubscript{RF1}$ represents the RF channel gain between the transmitter and the power cable, $H\textsubscript{PLC}$ denotes the channel gain on the power cable, and $H\textsubscript{RF2}$ expresses the channel gain between the power cable and the receiver. It is assumed that the signals radiated by the power cable do not significantly couple back into the PLC channel, as their strenght is weak \cite{5350425}. The system is analyzed within a mining environment characterized by a lossy dielectric ground (\(\varepsilon_r = 10\), \(\sigma = 0.01 \, \text{S/m}\)) and multi-path effects.

\subsection{Indoor Power Cable Radiation Theory}
\begin{figure}
    \centering
    \includegraphics[width=0.8\linewidth]{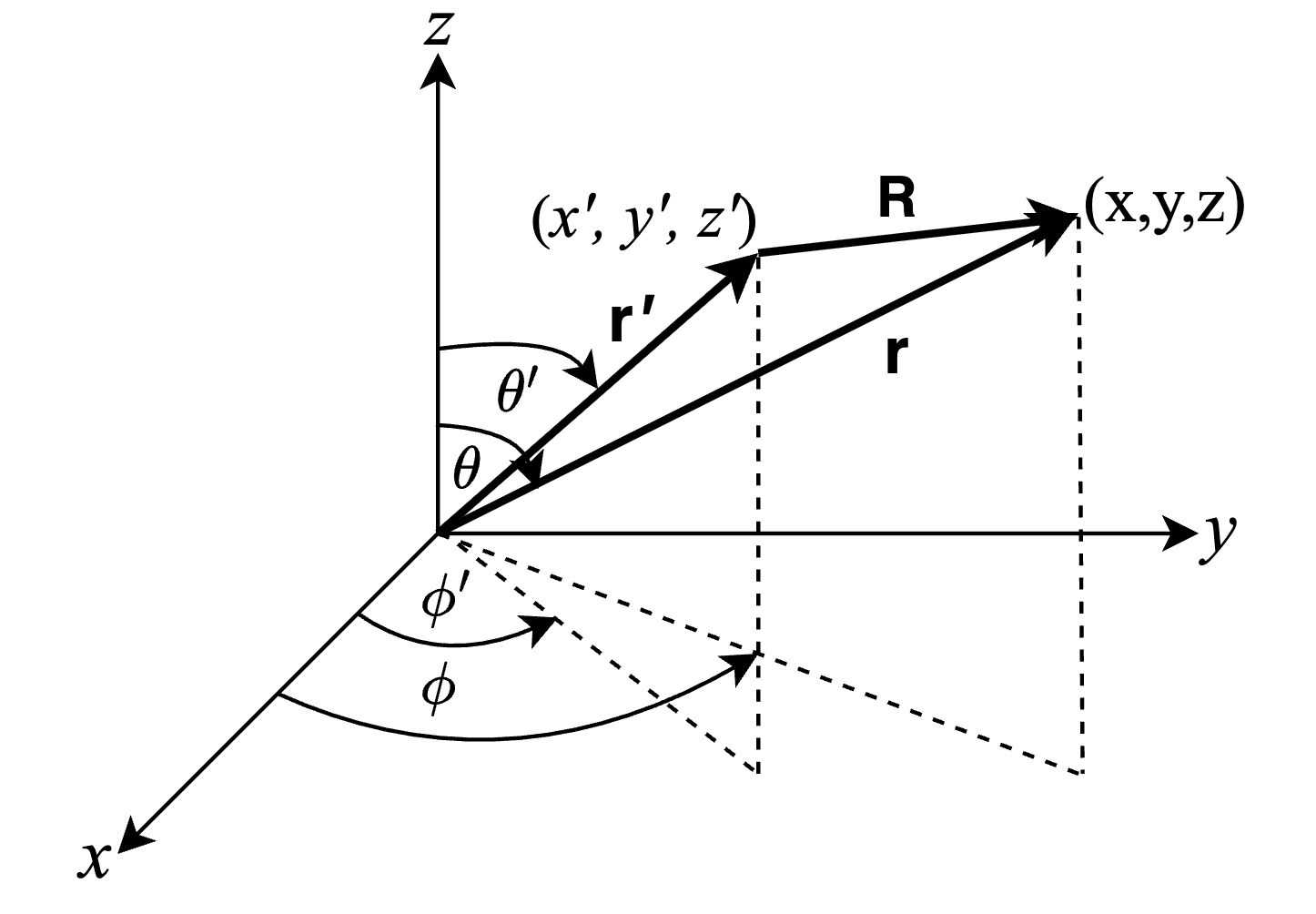}
    \caption{Coordinate system for radiated fields.}
    \label{fig:cordinate_system_for_radiated_fields}
\end{figure}

\noindent As in any antenna, indoor power cable radiation is governed by Maxwell’s equations, which describe the relationship between electric (\(\mathbf{E}\)) and magnetic (\(\mathbf{B}\)) fields \cite{Balanis:2016}. The Ampere-Maxwell law, in time-harmonic form, can be expressed as 
\(\nabla \times \mathbf{B} = \mathbf{J} + j \omega \mathbf{D_e}\) \cite{Balanis:2016},
\noindent where \(\mathbf{J}\) represents the density of the conduction current, \(\mathbf{D_e} = \varepsilon \mathbf{E}\) denotes the electric displacement, \(\omega = 2\pi f\) expresses the angular frequency, and \(\varepsilon\) stands for permittivity. Figure \ref{fig:cordinate_system_for_radiated_fields}, adapted from \cite{Balanis:2016}, shows the coordinate system for radiated fields and helps to find an expression for a wire carrying an \gls{rf} current \(I(z)\). This expression of the magnetic vector potential \(\mathbf{A}\) can be given by \cite{Balanis:2016}

\begin{equation}
\mathbf{A}(\mathbf{r}) = \frac{\mu_0}{4\pi} \int_C \frac{I(z') e^{-j k |\mathbf{r} - \mathbf{r}'|}}{|\mathbf{r} - \mathbf{r}'|} dz',
\label{eq:vector_potential}
\end{equation}

\noindent where \(\mu_0 = 4\pi \times 10^{-7} \, \text{H/m}\) represents the free-space permeability, \(k = \frac{2\pi}{\lambda}\) expresses the wave number,  \(\lambda = \frac{c}{f}\) indicates the wavelength, and \(c = 3 \times 10^8 \, \text{m/s}\), refers to the speed of light in a vacuum. The parameter \(|\mathbf{r} - \mathbf{r}'|\) designates the resultant distance between the three-dimensional vectors \(\mathbf{r}\) and \(\mathbf{r}'\). The electric field can be derived as
\(\mathbf{E} = -j \omega \mathbf{A} - \nabla \phi\), where \(\phi\) denotes the scalar potential, solved via the Lorenz gauge. \subsubsection{Radiation Pattern}
It describes the angular distribution of the radiated power. For a long-wire antenna, the far-field also known as the Fraunhofer region of the electric field in spherical coordinates can be given as \cite{Balanis:2016}

\begin{equation}
E_\theta = j \eta \frac{k I_0 e^{-j k \mathbf{r}}}{4\pi \mathbf{r}} \sin \theta \int_{-L_w/2}^{L_w/2} e^{j k z \cos \theta} dz,
\label{eq:far_field}
\end{equation}

\noindent where \(\eta = 120\pi \, \Omega\) represents the free-space impedance, \(I_0\) denotes the peak current, \(L_w\) expresses the wire length, \(\mathbf{r}\) stands for the radian distance to the observation point, and \(\theta\) defines the angle from the wire axis. The normalized radiation pattern can be evaluated as

\begin{equation}
F(\theta) = \frac{\sin\left(\frac{\pi n \cos \theta}{2}\right)}{\sin \theta}, \quad n = \frac{L_w}{\lambda},
\label{eq:norm_pattern}
\end{equation}

\noindent where \(n\) expresses the electrical length, which is mostly dependent on the physical length of the wire, the operating frequency, and the type of cable used. The pattern exhibits multiple lobes, with the number of lobes approximately equal to \(2n\), due to standing waves along the wire.\subsubsection{Directivity}
\noindent This parameter quantifies the concentration of the radiated power in a specific direction. It is defined as \cite{Balanis:2016}

\begin{equation}
D(\theta, \phi) = \frac{4\pi U(\theta, \phi)}{P_{\text{rad}}},
\label{eq:directivity}
\end{equation}

\noindent where \(U(\theta, \phi) = \frac{\mathbf{r}^2 |E_\theta|^2}{2\eta}\) represents the radiation intensity, and \(P_{\text{rad}}\) the total radiated power. For a long-wire antenna, the maximum directivity increases with the electrical length and is approximated as 
\(D_{\text{max}} \approx 2 \log_{10} (2n) \, \text{dBi}\). The main lobe’s angle shifts toward end-fire directions (\(\theta \approx 0^\circ, 180^\circ\)) as \(n\) increases with
\begin{equation}
\theta_{\text{max}} \approx \cos^{-1} \left( \sqrt{1 - \frac{0.371}{n}} \right).
\label{eq:theta_max}
\end{equation}

\subsubsection{Input Impedance and Radiation Resistance}
\noindent The antenna input impedance, \(Z_{\text{in}} = R_r + R_{\text{ohm}} + jX\), determines the efficiency of power transfer, where \(R_r\) represents the resistance to radiation, \(R_{\text{ohm}}\) the ohmic loss, and \(X\) the reactance. For a long-wire antenna, the radiation resistance is approximated as 
\(R_r \approx 120 \left( \ln\left(\varsigma\right) - 1 \right)\)\cite{Balanis:2016}, where \(\varsigma= \frac{2L_w}{d}\), and \(d\) denotes the wire diameter. The impedance varies with frequency and length, exhibiting oscillatory behavior due to standing waves. Matching \(Z_{\text{in}}\) to the feed line maximizes power transfer \cite{seybold2005introduction}.\\

\subsubsection{Indoor Long-Wire Antenna Modeling}
\noindent A long-wire antenna, defined as having a length \(L_w \geq \lambda\), supports traveling or standing waves depending on termination \cite{arrl:2019}. In this study, the power cable is modeled as a traveling wave antenna, terminated with its characteristic impedance \(Z_0\), which prevents reflections according to \cite{Krause:1950}. The retarded current can be expressed as 
\(I(z, t) = I_m \sin \omega \left( t - \frac{r}{c} - \frac{z}{v} \right)\), where \(v = p c\) is the wave velocity in the wire, and \(p\) denotes the relative phase velocity. This model captures the directional radiation properties essential for \gls{cplc}. In the case under investigation, the power cable is treated as a random wire, horizontally placed on the ground, excited at one end by an \gls{rf} source and terminated at the other with \(Z_0\). The cable’s geometry influences its radiation properties. The far-field pattern, directivity, and impedance are computed using the \gls{mom} to solve the electric field integral equation as \(E_z^{\text{inc}}(\mathbf{r}) = -j \omega A_z(\mathbf{r}) - \frac{\partial \phi(\mathbf{r})}{\partial z}\)\cite{Balanis:2016}. The \gls{mom} discretizes the wire into segments, solving for the current distribution, \(I(z)\), through the impedance matrix, \([Z][I] = [V]\). According to \cite{seybold2005introduction}, ground effects can be modeled using image theory, adjusting the radiation pattern for reflections from the lossy ground.

\subsection{\gls{cplc} System Model}
\noindent 

\subsubsection{PLC Channel Model}
\begin{figure}
    \centering
    \includegraphics[width=0.38\linewidth]{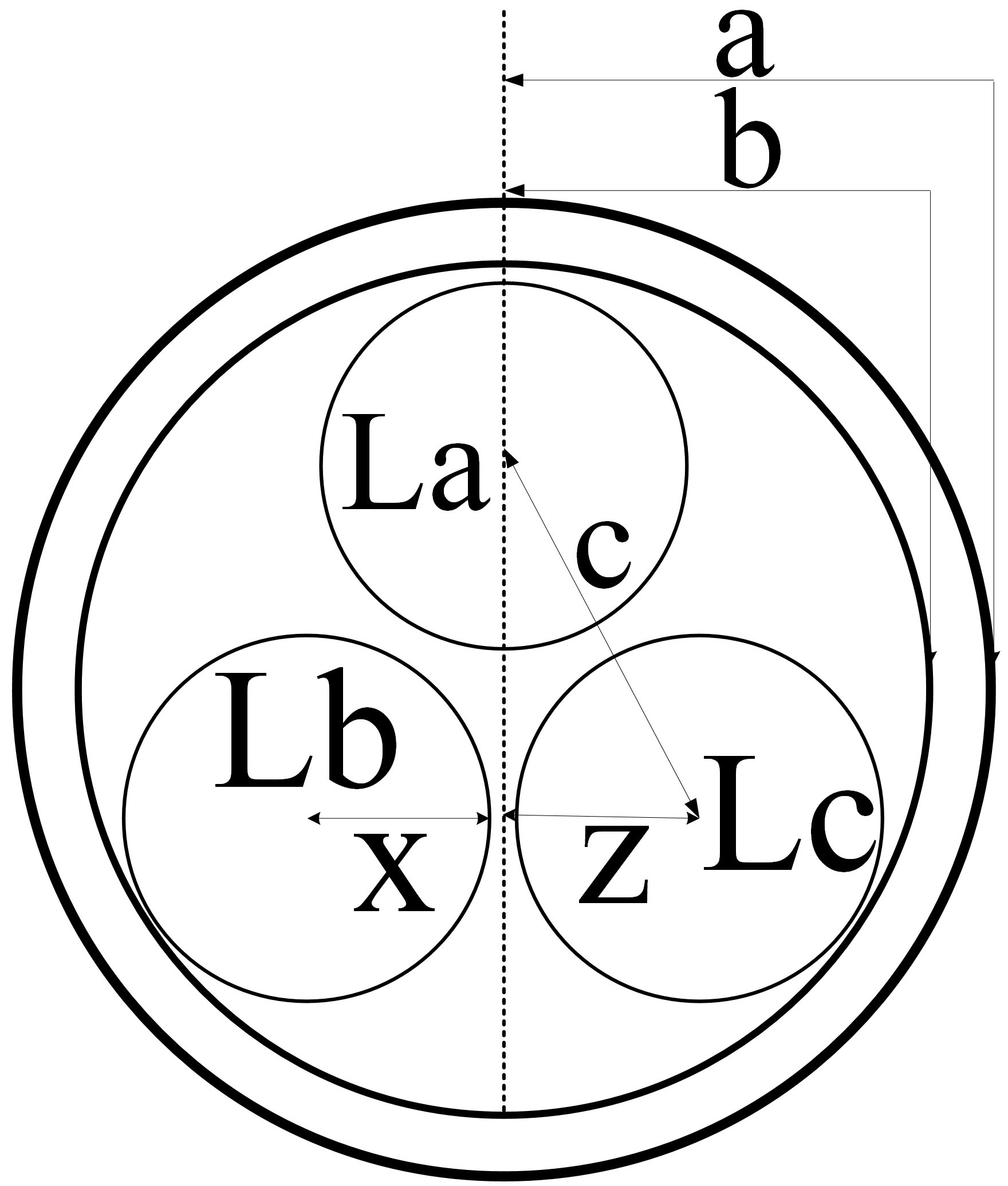}
    \caption{Three phase mining cable.}
    \label{fig:3_phase_cable}
\end{figure}

\noindent The PLC channel serves as the primary medium utilized in a \gls{cplc} system \cite{igboamalu2021contactless}. The cable analyzed in this work is a three-phase mining cable, featuring a central copper screen and reinforced with steel armour for mechanical protection, as illustrated in Fig.~\ref{fig:3_phase_cable}. The cable consists of three uniformly spaced phase conductors designated as, La, Lb, and Lc, each made of stranded copper and corresponding to the three phases of a standard three-phase power system, insulated with Ethylene Propylene Rubber. The frequency response of the PLC channel can be given as \cite{zimmermann2002multipath,1372925} 

\begin{equation}
        H_{\text{PLC}} = \sum_{i=1}^{N} g_i \cdot \Lambda(f,\ell_i) \cdot e^{-j 2\pi f \tau_i},
        \label{eq:plc}
\end{equation}

\noindent where $g\textsubscript{i}$ is the weighting factor of a specific propagation path, which is a function of the transmission and reflection factors. The weighting factor decreases with the increase in the signal propagation paths. Extended measurements showed that the weighting factors can be simplified, as they do not need to vary with frequency and can often be treated as real numbers. From a multipath perspective, $g\textsubscript{i}$ represents the strength or contribution of each path \cite{zimmermann2002multipath}. \textit{N} denotes the number of propagation paths, \textit{f} is the frequency, and \(\tau_i\) stands for the propagation delayed arrival time of the specific path. \textit{$\Lambda(f,\ell)$} represents the attenuation portion of the frequency response for a signal traveling a distance $\ell$ at a frequency $f$. \( \tau_i = (\ell_i \epsilon_r/c) = \ell_i/v_p \), where \( \varepsilon_r \) is the relative dielectric constant of the cable insulation, \textit{$v_p$} denotes the phase velocity.  \textit{$\Lambda(f,\ell)$} models how the signal is weakened as it propagates through the power line cable and can be defined as
\( \Lambda(f,\ell) = e^{-\alpha(f)\ell}\), where \(\alpha(f)\) is a frequency-dependent attenuation factor. \(\alpha(f)\) can be determined by the complex propagation constant \(\gamma\) \cite{zimmermann2002multipath}, and evaluated as
\begin{equation}
        \gamma = \alpha(f) + j\beta(f) = \sqrt{(R + j2\pi f\mathcal{L}) + (G + j2\pi fC)}.
\end{equation}

\noindent The complex propagation constant is governed by the primary parameters of the power  cable, namely resistance (\textit{$R$}), inductance (\textit{$\mathcal{L}$}), capacitance (\textit{$C$}), and conductance (\textit{$G$}). These parameters exhibit frequency-dependent behaviors. Substituting the expression for \textit{$\Lambda(f,\ell)$} into the definition of $H\textsubscript{PLC}$, we obtain
\begin{equation}
       H_\text{PLC} = \sum_{i=1}^{N} g_i e^{-\alpha(f) \ell_i} e^{-j 2\pi f /\tau_i}.
       \label{eq:plc_comp}
\end{equation}

\noindent As discussed in section 4 of the indoor power cable radiation theory, the power cable is modeled as a traveling wave antenna terminated with \textit{$Z_0$}, defined as \cite{tsuzuki2002estimation}
\begin{equation}
       Z_\text{0} = \sqrt{\frac{R +j\omega \mathcal{L}}{G + j\omega C}}.
       \label{eq:z0}
\end{equation}

\subsubsection{RF Channel}
 As \gls{rf} signals propagate through the environment, they are subject to reflection, diffraction, and scattering due to interactions with physical objects and atmospheric particles. The extent of these effects depends on factors such as the number, size, material, and position of obstacles between the transmitter and receiver, as well as the transmission distance. The RF channel gain can be given as  \cite{231342, igboamalu2021contactless}
\begin{equation}
       H_\text{RF} = \sum_{i=1}^{M} a_i \cdot e^{-j2\pi f \tau_n}.
       \label{eq:rfc}
\end{equation}
\noindent Here, \textit{$a_i$}  and \textit{$\tau_n$}  represent the complex amplitude and the propogation delay of the \textit{$m_{\text{th}}$} wave, while \textit{$M$} denotes the total number of waves.
\newline
\subsubsection{\gls{cplc} Channel Model}
The complete \gls{cplc} channel model is derived by integrating \eqref{eq:plc_comp} and~\eqref{eq:rfc}. The resulting formulation constitutes the proposed model, which represents the \gls{cplc} channel, $H_\textsubscript{CPLC}$, can be expressed as 
\begin{equation}
       H_\text{CPLC} = \left(\sum_{i=1}^{N} g_i e^{-\alpha(f) \ell_i} e^{-j {2\pi f}{\tau_i}} \right)\left(\sum_{n=1}^{M}  a_n e^{-j 2\pi f \tau_n} \right)^2.
       \label{eq:cplc}
\end{equation}

\noindent This model assumes that $H\textsubscript{RF1}$ and $H\textsubscript{RF2}$ are equal and that the coupling between $H\textsubscript{PLC}$, $H\textsubscript{RF1}$ and $H\textsubscript{RF2}$ is perfectly efficient.

\section{Simulation Results and Analysis}

\begin{figure}
    \centering
    \includegraphics[width=0.85\linewidth]{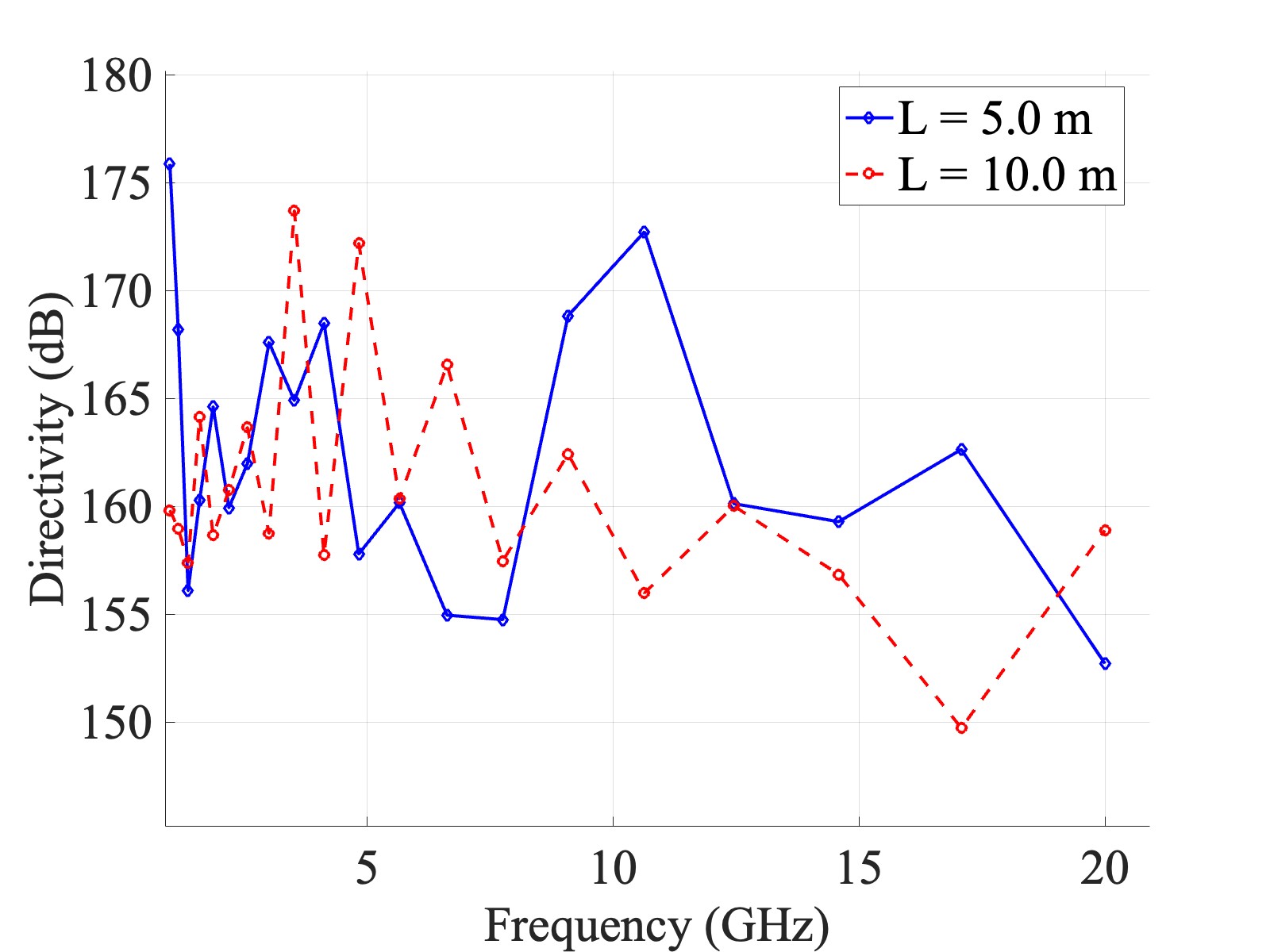}
    \caption{Directivity changes with change in frequency.}
    \label{fig:DvsF_5m_10m}
\end{figure}

\begin{figure}
    \centering
    \includegraphics[width=0.8\linewidth]{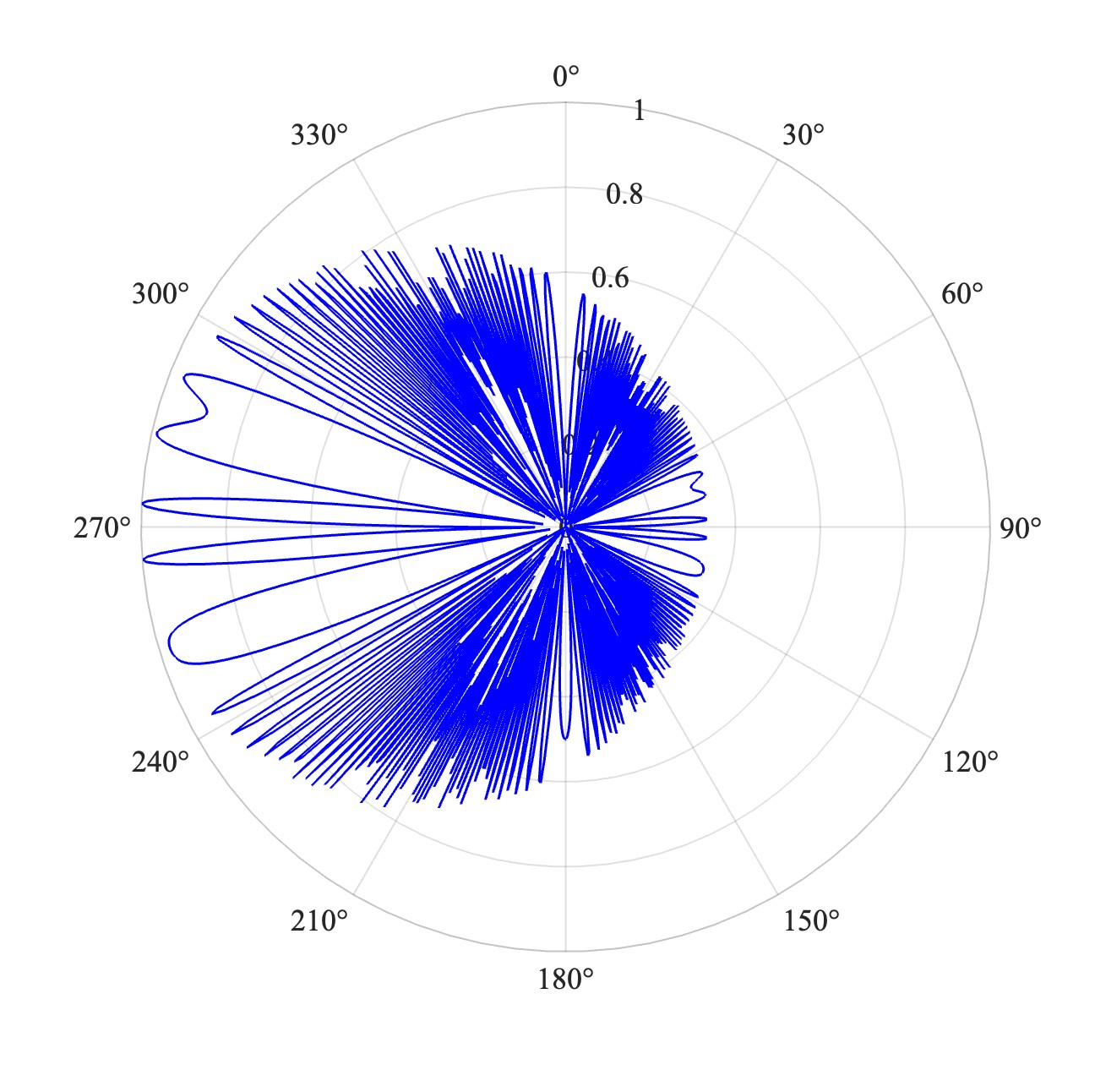}
    \caption{Radiation pattern at 10 GHz.}
    \label{fig:polar_plot2_10GHz_5m}
\end{figure}

\noindent Figure \ref{fig:DvsF_5m_10m} shows the changes in directivity as the frequency varies at a given length.  In Fig. \ref{fig:polar_plot2_10GHz_5m} it is observed how the gain changes with angle, based on a simulated radiation pattern for a 5 m long-wire antenna operating at 10 GHz. Figure \ref{fig:polar_plot2_10GHz_5m} also shows the multiple main lobes that have gains greater than 80\% of the maximum power on the normalized scale at angles between 240\degree ~and 300\degree, in the azimuthal plane, oriented according to the cable lying horizontally on the ground in an indoor mining tunnel. The multiple lobes observed typically arise due to the electrical length of the wire relative to the operating wavelength. When a wire antenna is much longer than a quarter wavelength, it no longer behaves like a simple monopole. Instead, it starts to exhibit standing wave patterns, leading to multiple lobes in the radiation pattern. These lobes are areas of constructive interference where the radiated signal is stronger. The behavior of the indoor long-wire antenna depends heavily on the current distribution, and as such, in a short monopole, the current distribution is roughly triangular, peaking at the feed-point. However, in a long-wire, multiple current maxima and minima form along the wire, creating multiple radiation zones. Each of these zones contributes to a lobe in the far-field pattern. 

\begin{table}[b]
  \centering
  \caption{Cable parameters.}
  \label{tab:parameters}
  \begin{tabular}{|c|c|c|c|}
    \hline
    \textbf{Parameter} & \textbf{Value (mm)} & \textbf{Parameter} & \textbf{Value} \\
    \hline
    a           & 37.45           & $\epsilon_r$              & 2.25 \\
    b           & 29.65           & $\mu_c$                   & 0.999994 \\
    c           & 13.51           & $\sigma_c$ (S/m)          & $5.8 \times 10^7$ \\
    x           & 3.6             & $\sigma_{\text{epr}}$ (S/m)     & $1 \times 10^{-13}$ \\
    z           & 33.85           & $\mu_a$                   & 1 \\
    \hline
  \end{tabular}
\end{table}

\begin{figure}
    \centering
    \includegraphics[width=0.8\linewidth]{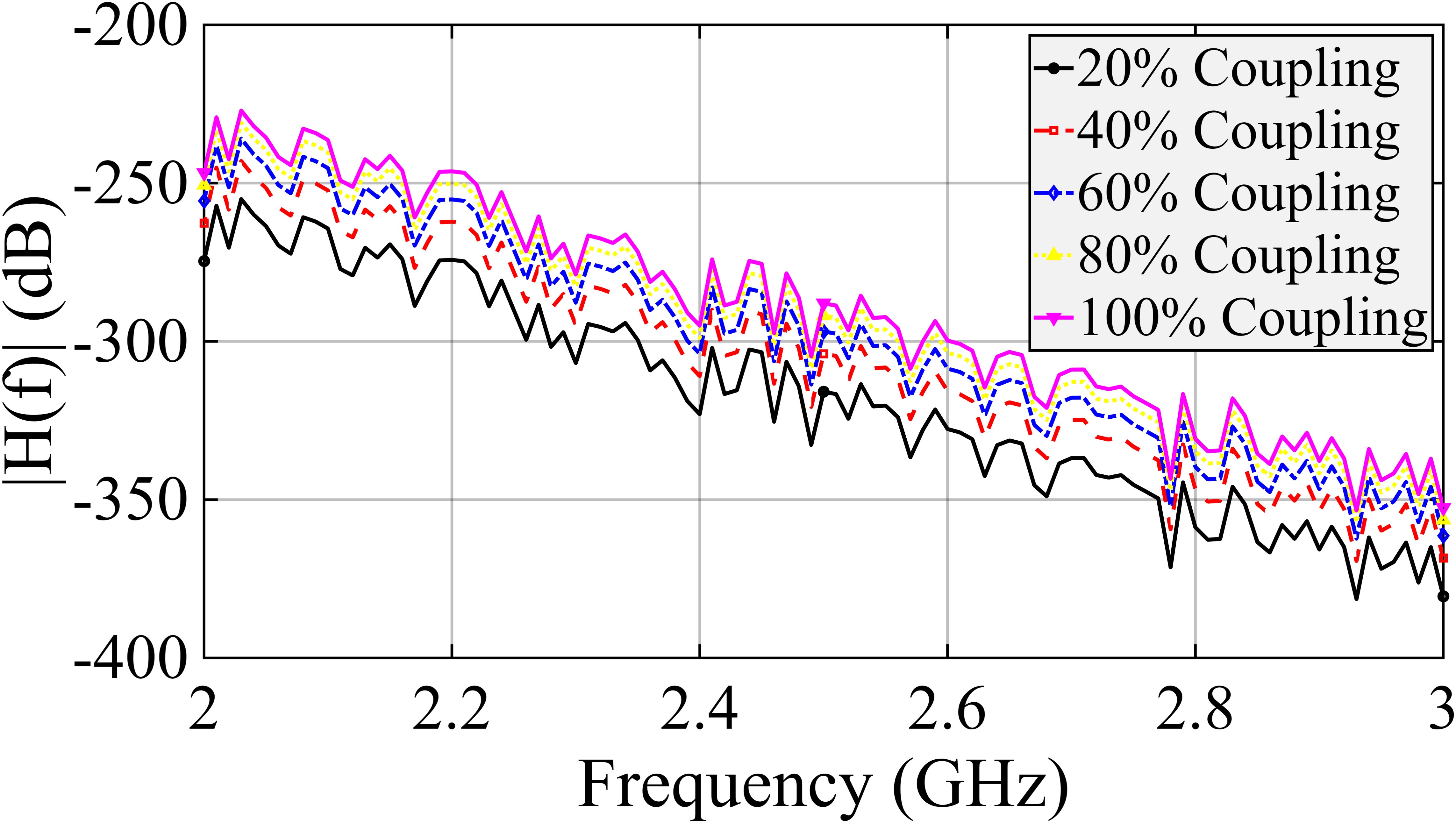}
    \caption{CPLC channels with varying coupling efficiency.}
    \label{fig:fig_varying_eps}
\end{figure}

\begin{figure}
    \centering
    \includegraphics[width=0.8\linewidth]{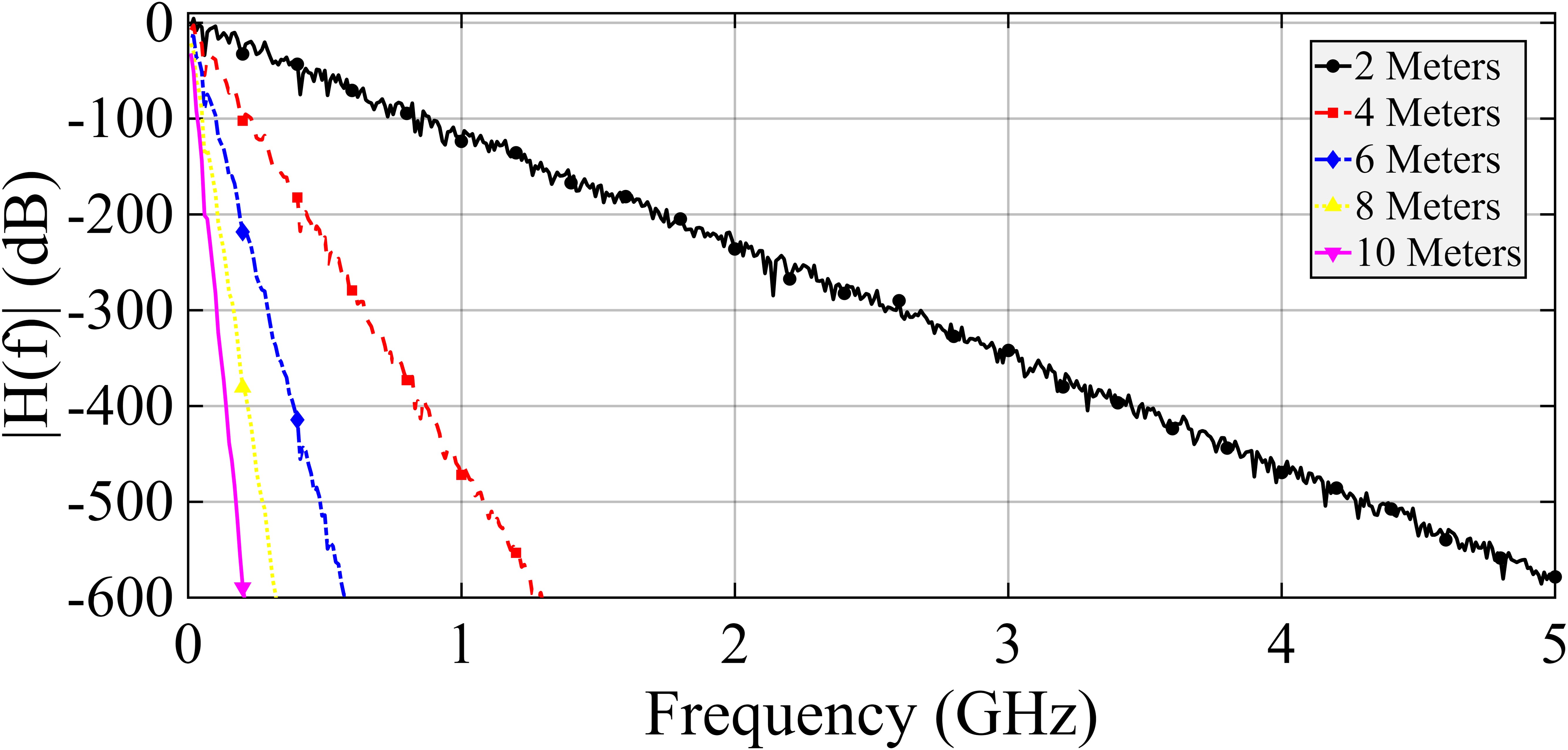}
    \caption{CPLC channels with varying distances.}
    \label{fig:fig_varying_distanace}
\end{figure}

\begin{figure}
    \centering
    \includegraphics[width=0.9\linewidth]{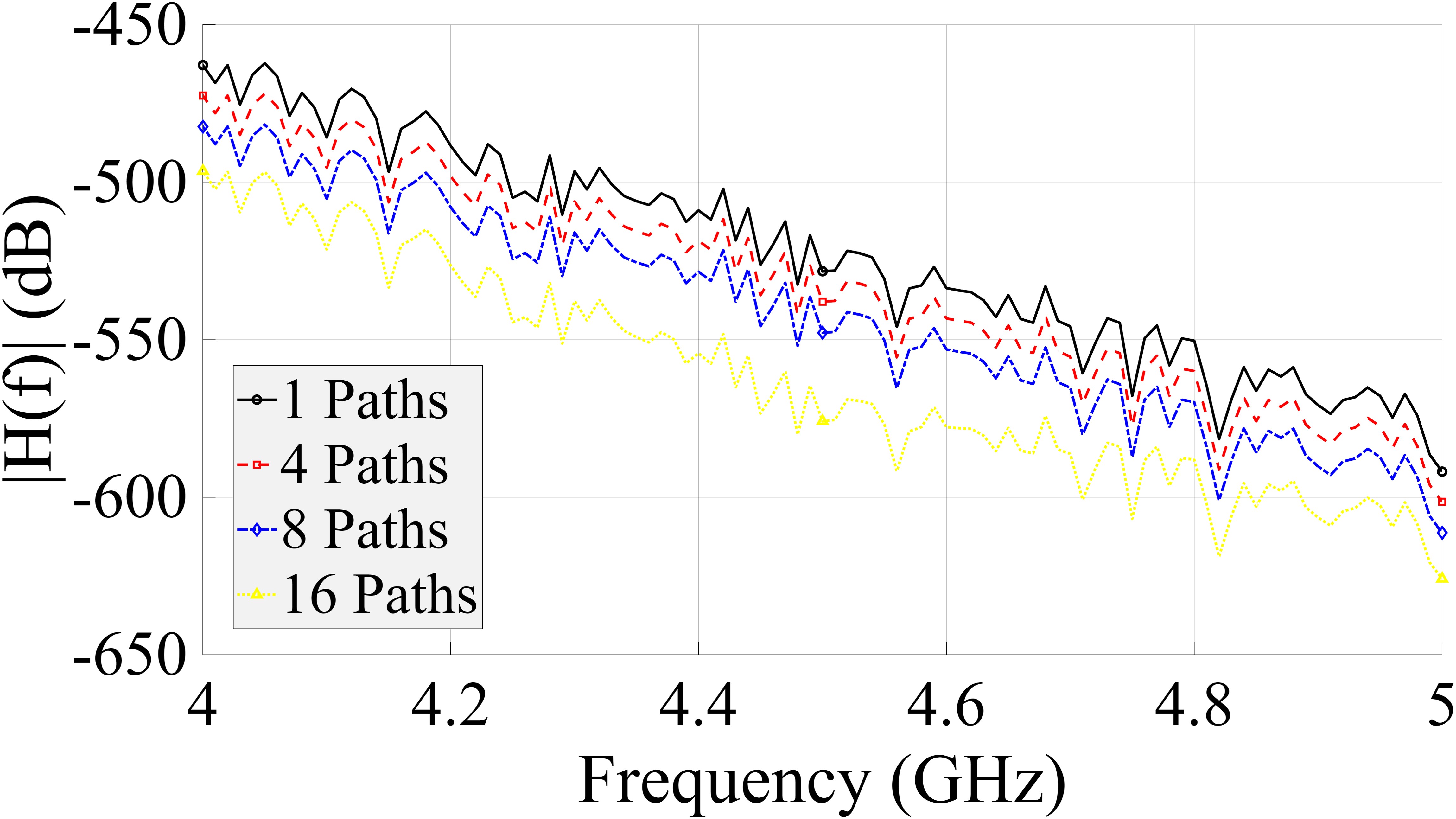}
    \caption{CPLC channels with varying propagation paths.}
    \label{fig:fig_varying_paths}
\end{figure}
\noindent Equation \eqref{eq:cplc} is implemented to characterize the \gls{cplc}  channel. Table~\ref{tab:parameters} lists the parameters of the selected cable. The Rician distribution is considered to generate the $H\textsubscript{RF1}$ and $H\textsubscript{RF2}$ channel gains. The simulation assumed a \gls{los} distance of 3 m between the power line cable and the receiving RF device. The simulations are conducted across a frequency range from 0 Hz to 5 GHz, with 10 MHz intervals. In Fig. \ref{fig:fig_varying_eps}, the coupling efficiency between the \gls{plc} and \gls{rf} channels is varied, while the \gls{plc} cable length is fixed at 2 m and a single propagation path is assumed. This figure indicates that a lower coupling efficiency between 
$H\textsubscript{PLC}$ and $H\textsubscript{RF1}$, $H\textsubscript{PLC}$ and $H\textsubscript{RF2}$ results in greater signal attenuation within the system. Since coupling occurs at two separate interfaces, the total attenuation compounds multiplicatively. Therefore, optimizing the coupling efficiency at each interface is critical to minimizing overall signal degradation.
In Fig. \ref{fig:fig_varying_distanace}, the \gls{plc} cable length is varied under the assumption of a single propagation path and a coupling efficiency of 100\%. This figure indicates that the attenuation of the overall system increases with the increase in cable length.
In Fig. \ref{fig:fig_varying_paths}, the \gls{plc} cable length is fixed at 2 m, and the coupling efficiency is maintained at 100\%, while the number of propagation paths within the \gls{plc} channel is varied. Since the distance between the power cable and the receiving \gls{rf} device is relatively short, the parameters of the \gls{rf} channel are not varied. Figure \ref{fig:fig_varying_paths} also illustrates that an increase in the number of propagation paths can introduce additional reflection losses, which may lead to greater attenuation of the input signal. 

\section{Conclusion}
\noindent This research developed a framework for using antenna radiation theory in CPLC systems for underground mining. It confirmed that mining power cables can act as long-wire antennas, with their length having a significant impact on signal directivity. The proposed model captured key effects like frequency-dependent attenuation, demonstrating that longer cables produce multi-lobed radiation patterns. This CPLC system could improve mobility and leverage existing infrastructure, overcoming limitations of current systems. Future work should investigate the mutual coupling effects of multiple active wires.

\bibliographystyle{IEEEtran}
\bibliography{references_IEEE}
\end{document}